\documentclass[pra,11pt,onecolumn, showkeys,showpacs]{revtex4}

\usepackage{times}

\begin{document}

\title{Comment on "Cryptanalysis and improvement of multiparty quantum secret sharing schemes"}

\author{Gan Gao\footnote{Corresponding author. E-mail:
gaogan0556@163.com}}

\affiliation{Department of Electrical Engineering, Tongling
University, Tongling 244000, China}

\date{\today}

\begin{abstract}

We show that, using Wang et al. attack [T.-y. Wang, Q.-y. Wen, F.
Gao, S. Lin, F.-c. Zhu, Phys. Lett. A $373$ (2008) 65], the first
agent and the last agent cannot eavesdrop all the secret messages in
Zhang et al. QSSCM scheme [Z.-j. Zhang, G. Gao, X. Wang, L.-f Han,
S.-h. Shi, Opt. Commun. $269$ (2007) 418]. In some sense, Wang et
al. attack is unsuccessful for Zhang et al. QSSCM scheme.
\end{abstract}



\maketitle

Recently, using Bell states and local operations, Zhang {\it et al.}
[1] proposed an interesting scheme of multiparty quantum secret
sharing of classical messages (QSSCM). However, it is a slight pity
that this scheme has a drawback of security, which has been pointed
out by Lin {\it et al.} [2]. Lin {\it et al.} showed that Charlie
may solely eavesdrop half of the secret messages without introducing
any error. Several months later, Wang {\it et al.} [3] proposed
another attack on Zhang {\it et al.} QSSCM scheme [1] (By the way,
there are four same authors in Ref. [2] and Ref. [3]). In their
attack, it was shown that the first agent and the last agent may
collaborate to eavesdrop all the secret messages without the helps
of other agents. Since the quantity of eavesdropped messages is
twice as much as that in Lin {\it et al.} attack, Wang {\it et al.}
claimed that their attack is stronger than Lin {\it et al.}'s.
However, this is not fact. In this paper, we will show that Wang
{\it et al.} attack cannot eavesdrop all the secret messages, and it
cannot eavesdrop one half until a legal mode is added. This means
their attack is not stronger than Lin {\it et al.}'s, and their
statement in the abstract of [3], {\it "We further show the first
agent and the last agent can obtain all the secret without
introducing any error in Zhang's et al. multiparty QSSCM scheme by a
special attack with quantum teleportation"}, is incorrect. Next, let
us simply review Lin {\it et al.}'s and Wang {\it et al.}'s attacks
as
follows.\\

{\it Lin {\it et al.} attack}$-$When Charlie has received photon
$t$, he randomly chooses one of the following two procedures. (a)
The legal mode. Charlie performs a $H$ operation on the photon and
then acts according to the legal process. (b) The attack mode.
Charlie sends a fake photon $t'$ from
$\psi_{h't'}^{-}=(|0\rangle|1\rangle-|1\rangle|0\rangle)_{h't'}/\sqrt{2}$
to Alice and retain photon $h'$. After Alice received photon $t'$,
in terms of the mode chosen by Alice, Charlie applies corresponding
operation as follows. In the message mode, Alice performs a unitary
operation on photon $t'$ to encodes her secret message. When Alice
sends photon $t'$ to Bob, Charlie intercepts it and makes a Bell
state measurement on photons $h'$ and $t'$. So he can deduce Alice's
secret message. In the control mode, when Alice asks Charlie to
declare his operation, Charlie makes a Bell state measurement on
photons $h'$ and $t$. Then, according to his measurement outcome,
Charlie announces a fake information on his unitary operation. Since
Alice and Bob will get the expected result during the
eavesdropping-check procedure, Alice cannot find Charlie's cheating.

Lin {\it et al.} attack shows that, after receiving the photon $t$,
Charlie randomly chooses one of two procedures: the legal mode and
the attack mode. Here, although the probability that each mode is
chosen isn't definitely given, they should be both 50\%. Next, let
us analyze why Lin {\it et al.} set up the legal mode in which
Charlie performs only $H$ operation. We see that, in the control
mode, the fake operation that Charlie may announce to Alice is
limited in the four unitary operations: $I$, $\sigma_{x}$,
$\sigma_{y}$, $\sigma_{z}$. If he is always publishing the four
unitary operations, this will cause Alice's suspicion, that is, she
may think Charlie is eavesdropping. This is because, in Zhang {\it
et al.} QSSCM scheme [1], the probability that all the agents except
for Bob select to perform the $H$ operation is 50\%, which leads to
the probability that the $H$ operation is published should be also
50\% (By the way, the probability that the agents select each
operation is known by all the participants in Zhang {\it et al.}
QSSCM scheme, and the value of the probability should be fixed and
agreed in advance). In order to let $H$ operations appear in his
publishing operations, Charlie must have the legal mode. So it may
be confirmed that Lin {\it et al.} set up the legal mode is
reasonable. Since the legal mode exists, obviously, Charlie cannot
do replacing actions for all photons $t$, but choose that one half
are replaced with and the other half aren't. For instance, there are
100 photons $t$ in one turn, he randomly replaces 50 photons $t$
with 50 fake photons $t'$, and the other 50 ones can't be replaced
with. Relying on the replaced ones, Charlie can eavesdrop half of
Alice's secret messages.
\\

{\it Wang {\it et al.} attack}$-$When Zach receives photon $t$, he
prepares a four-qubit state $|g_{1}\rangle_{1234}$ and sends photon
4 in the $|g_{1}\rangle_{1234}$ to Alice, and sends photons $t$123
to Bob. When Alice switches to the control mode, Bob performs a $G$
state measurement on photons $ht$12. So the state of photons $ht$ is
teleported to photons $34$. According to the evolved equation of the
$"ht1234"$ system, Bob secretly tells Zach a fake unitary operation.
When Alice asks the agents to publish operations, Zach announces the
fake unitary operation to her. Since Alice will get the expected
result during the eavesdropping-check procedure, she cannot find
Bob's and Zach's deceiving. When Alice switches to the message mode,
she performs a unitary operation on photon 4 to encode her secret
messages, and then sends photon 4 to Bob. Bob and Zach perform a $G$
state measurement on photons 1234, and deduce all the secret
messages.

It is evident that the legal mode is not set up in Wang {\it et al.}
attack [3]. From beginning to end, Bob and Zach are always doing the
replacing action. That is, all photons $t$ are replaced with by them
during the secret sharing. When the control mode is switched into,
Bob performs a $G$ state measurement and teleports the state of
photons $ht$ to photons $34$, and secretly tells Zach a fake
operation according to the evolved equation of the $"ht1234"$
system. Note that, since the fake operation belongs to one of the
four unitary operations: $I$, $\sigma_{x}$, $\sigma_{y}$,
$\sigma_{z}$, the operation that Zach can announce to Alice must be
also limited in the four ones. However, a explanation has been given
in the above content, that is, the $H$ operation should appear with
the probability 50\% in Zach's announcing operations. If no, Alice
can {\it at least} judge that Zach is eavesdropping. Based on this
reason, Wang {\it et al.} attack may be regarded as a unsuccessful
attack in some sense. Of course, if only a legal mode similar to the
one in Lin {\it et al.} attack is added, their attack will become
feasible. But, after adding the legal mode, Wang {\it et al.} attack
has to face a fact that, it can eavesdrop only half of the secret
messages in Zhang {\it et al.} QSSCM scheme and is not stronger than
Lin {\it et al.} attack [2]. By the way, the main reason that Wang
{\it et al.} attack can also be accepted by {\it Physics Letters A}
after Lin
{\it et al.} attack has been published in {\it Optics Communications} is that, the late proposed attack is stronger than the early proposed one.\\

In summary, we have shown that, using Wang {\it et al.} attack, the
first agent and the last agent cannot eavesdrop all the secret
messages in Zhang {\it et al.} QSSCM scheme without introducing any
error, even their eavesdropping action may be detected by Alice,
because the $H$ operation doesn't appear in the last agent's
publishing operations. In order to make Wang {\it et al.} attack
feasible, a legal mode must be added. However, Wang {\it et al.}
attack added the legal mode can eavesdrop only half of the secret
messages as well as Lin {\it et al.} attack.
\\

\noindent {\bf Acknowledgements}\\

I thank my parents for their encouragements. \\

\noindent {\bf References}

\noindent[1] Z. J. Zhang, G. Gao, X. Wang, L. F. Han and S. H. Shi
 Opt. Commun.  $269$ (2007) 418.

\noindent[2] S. Lin, F. Gao, Q. Y. Wen, F. C. Zhu, Opt. Commun.
$281$ (2008) 4553.

\noindent[3] T. Y. Wang, Q. Y. Wen, F. Gao, S. Lin, F. C. Zhu, Phys.
Lett. A $373$ (2008)65.

\enddocument